# Nanofabricated media with negative permeability at visible frequencies

A. N. Grigorenko*, H. F. Gleeson*, Y. Zhang*, A. A. Firsov*†, I. Y. Khrushchev‡, J. Petrovic‡

*Department of Physics and Astronomy, University of Manchester, Manchester, M13 9PL, UK

†Institute of Microelectronics Technology, 142432 Chernogolovka, Russia

‡Department of Electronic Engineering, Aston University, Aston Triangle, Birmingham, B4 7ET, UK

**A great deal of attention has recently been attracted to a new type of smart materials, so-called left-handed media, which exhibit highly unusual electromagnetic properties and promise unique device applications, including a perfect lens [1-9]. Left-handed materials require negative permeability $\mu$, an extreme condition that so far was achieved only for microwave frequencies by using arrays of metal-wire resonators [1,4-13]. Extension of this approach from microwaves to visible light presents a formidable if not impossible challenge [12-17], as no material – natural or artificial – is known to exhibit any magnetism above terahertz frequencies. Here we report a nanofabricated medium made of electromagnetically coupled pairs of gold dots with geometry carefully designed at a 10-nm level. The medium exhibits strong magnetic response at visible-light frequencies, including bands with negative $\mu$. The magnetism arises due to the excitation of quadrupole plasmon resonances. Our approach shows for the first time the feasibility of magnetism at optical frequencies and paves a way towards magnetic and left-handed components for visible optics.**

The standard theory forbids any significant magnetism at optical frequencies. Indeed, the renowned reference in theoretical physics by Landau and Lifshitz explicitly reads [16] "there is certainly no meaning in using the magnetic susceptibility from optical frequencies onwards, and in discussion of such phenomena we must put $\mu$=1". This statement is strongly supported by experiment: magnetic susceptibility $\chi$ of all natural materials tails off at microwave frequencies [16]. Still, there may be a way to overcome the fundamental limitations. Pendry *et al* have recently suggested exploiting the inductive response of normal metals to obtain high-frequency magnetic response from structured nonmagnetic materials [10]. The idea was successfully implemented by using arrays of copper split-rings, and the man-made magnetism was demonstrated for frequencies up to 1 THz [11-13]. In this case, magnetism emerges due to collective motion of a large number of electrons (in essence, due to eddy currents), and theoretical arguments as, e.g., in ref. [16] – valid for individual electrons and atoms – no longer hold. It is tempting to exploit the idea further and create a visible-light version of magnetic and left-handed media. Unfortunately, the direct scaling [13] of the demonstrated microwave media to visible optics is impossible. This would require artificial structures with sizes down to 100nm – to match visible wavelengths – and critical features [10-13] controlled on the level of ≈10nm, which is hardly possible for the case of the split-ring geometry. Even more importantly, the scaling can fail on principle because of very different electromagnetic responses of materials to light and microwaves (e.g., it was predicted that inherent losses should limit the approach demonstrated in refs [11-13] to frequencies well below optical [12,17]).

We have achieved strong magnetic response at optical frequencies by using plasmon resonances excited in interacting metal structures of a sub-100nm size. Fig.1 shows an example of our devices and explains the basic idea behind the experiment. The studied structures were Au pillars microfabricated by high-resolution electron-beam lithography. Heights $h$ of Au pillars (85±5nm) and their diameters $d$≈100nm were chosen through numerical simulations so that the plasmon resonance characteristic to individual pillars appeared at red-light wavelengths, $\lambda \approx 670$nm. This resonance is clearly seen in the experimental spectra in Fig.2 (curve #1).

In addition to the reference structures, where individual Au pillars were well separated by distances $D \approx 0.5\mu$m, we have simultaneously fabricated samples consisting of exactly the same pillars but grouped in tightly spaced pairs. A number of different structures were studied with $d$ between 80 and 120nm and the separations $s$ between centres of adjacent pillars in the range 140 to 300nm and, i.e. the gap $s-d$ between the neighbouring pillars varied from 200nm down to just 20nm. At these distances, the electromagnetic interaction between neighbouring pillars becomes essential, and a plasmon resonance of an individual pillar splits into two for a pair of pillars (see Figs.1&2). These resonances are referred to as symmetric and antisymmetric, similar to the case of any classical or quantum system with two interacting parts and in agreement with the notations used for plasmon resonances in nanoparticles [18,19].

For the symmetric (also known as dipole) resonance (DR), electrons in neighbouring pillars move in phase, similar to the case of non-interacting pillars. The DR shifts to lower frequencies with increasing the interaction (decreasing $s$) (Fig.2c). The dipole mode is of secondary interest for the purpose of our report: It is the antisymmetric mode that gives rise to magnetic response. In the latter mode, electric charges in neighbouring pillars move in anti-phase so that the oscillating dipoles cancel each other leaving only a quadrupole response [18,19]. One can see in Fig.1c that the anti-phase movement of electrons along the *z*-axis effectively results in a current loop in the *z-x* plane. This high-frequency electric current generates a non-negligible magnetic field in the *y*-direction. Note a similarity between the double-pillar geometry and the split-ring resonators used in negative-$\mu$ microwave optics. Indeed, a pair of pillars can be thought of as a ring with two slits at the opposite sides. Such changes in the geometry of resonators when moving from microwaves to optics resemble the changes known for lasing techniques, where complex closed resonators used in masers were later replaced in lasers by a pair of mirrors [20,21].

The antisymmetric (quadrupole) resonance (QR) was observed in our experiments in the green-light range ($\lambda \approx$450-550nm) and only for structures with $s-d$ <100nm. The critical signature of this resonance is its shift to higher frequencies with decreasing $s$, directly opposite to the behaviour of the dipole resonance (Fig.2c). Using analogies with negative-$\mu$ microwave optics, one can expect a strong dependence of excitation efficiency of this resonance on light polarization [11]. The insets in Fig.3a show that this is also the case for our optical media. If illuminated by perpendicular white light, our structures look amber for one polarization (electric field along the x-axis), similar in colour to the media consisting of individual pillars or pairs with large $s$. This colour comes from DR. In contrast, for the perpendicular polarization, the structures usually look green due to a large contribution from the QR. The full spectral dependence of the reflectance as a function of the wavelength $\lambda$ and the polarization angle $\theta$ is plotted in Fig.3b.

The observed optical spectra are well described by the standard dispersion theory [22], which has also been used for analysis of microwave spectra of negative-$\mu$ materials. The DR is caused by an oscillating electric moment and, accordingly, contributes to permittivity as

$$\Delta\varepsilon(\lambda) = \frac{f_s \lambda^2}{\lambda^2 - \lambda_s^2 - i\lambda\Delta\lambda_s},$$

where $\lambda_s$ is the wavelength of the symmetric resonance and $\Delta\lambda_s$ its half-width. $f_s$ is an effective oscillator strength, which is proportional to the density of pillars and an efficiency of excitation of the resonance by incident light. There is no net magnetic moment attributable to DR. In contrast, for the quadrupole mode, the total dipole moment of a pillar pair is zero and, hence, QR's contribution to permittivity is zero. The electromagnetic response at the QR appears due to an oscillating magnetic moment, which results in "Pendry-type" permeability [12,13]

$$\mu(\lambda) = 1 + \frac{f_a \lambda_a^2}{\lambda^2 - \lambda_a^2 - i\lambda \Delta\lambda_a},$$

where $\lambda_a$, $\Delta\lambda_a$ and $f_a$ are the same notations as above but for the antisymmetric resonance. Note that non-zero excitation efficiency of this resonance requires a symmetry breaking, which was achieved by making pillars non-cylindrical (see Fig.3 caption). To calculate the sample reflectance, we combined these dispersion relations with Fresnel's coefficients for thin anisotropic films [22-24]. The solid lines in Fig.2a show results of our calculations. For examples, in the case of sample #3, our analysis yields $\Delta\lambda_s$=170nm and $\Delta\lambda_a$=57nm, $f_s$=0.17 and $f_a$=0.05. Note that the QR is noticeably narrower than the DR ($\Delta\lambda_a < \Delta\lambda_s$). We attribute this to the cancellation of radiative (dipole) losses in the quadrupole mode. According to the above analysis, magnetic susceptibility $\chi$ of sample #3 reaches a minimum value of about −0.23 near the QR. For sample #4 in Fig.3a, the diamagnetic response is larger, and $\chi$ reaches −0.5 at $\lambda_a$ ≈470nm. At the same wavelength, the contribution from DR also gives rise to a large negative permittivity ε ≈-0.98.

Magnetic response in the plasmon media can be increased by increasing the density of pillars. However, in this case, optical spectra become rather complicated, because of the interaction between non-nearest pillars. Instead, in order to increase $\chi$, we used the fact that QRs in nanoparticles [26,27] can be enhanced by placing them in a medium with a high refractive index $n$. To this end, we covered our samples with a thin (<100nm) layer of glycerine. This led to a significant increase in the strength of QRs (by a factor of 2 to 3), which was accompanied by their red shift by ≈50nm (see Fig.3a). The observed shift is in agreement with the amount of red shifts reported for QRs in nanoparticles [26]. The increased strength of QRs led to a proportional increase in $|\chi|$ so that negative values of $\mu$ were achieved routinely. For example, after covering sample

#4 with glycerine (Fig.3a), $\chi$ was found to be $\approx$–1.3 at the resonance (i.e. $\mu$=-0.3). Although some of our structures exhibited negative $\mu$ and $\varepsilon$ within the same range of $\lambda$ ($\varepsilon$=-0.7 for the sample #4 at the QR resonance), $\mu$ also had a large imaginary component ($\mu''\approx1i$ at the resonance), which so far did not allow the demonstration of the negative refraction. But there are no principle limitations for further enhancement of $\mu$ by using denser arrays or by improving the excitation efficiency in the quadrupole mode by optimising the design of plasmon resonators (e.g., by making pillars more conic). The latter seems to be the most viable way towards practical left-handed media at visible frequencies.

In our experiments, we have also obtained an independent proof for magnetism at optical frequencies, beyond the discussed comparison between experimental and theoretical spectra (note that such comparison is widely accepted as standard - see, e.g., [13-17]). It is the observation of impedance matching that allowed us to confirm directly and qualitatively that our structures had large $\chi$ at green-light wavelengths. Figure 4 explains the impedance matching phenomenon. The effect is characterised by the total suppression of the reflection from an interface between two media with different refraction indices $n=(\varepsilon\mu)^{1/2}$ but the same impedance values $Z=(\varepsilon/\mu)^{1/2}$. The impedance matching is well known in the physics of electrical circuits and antennas but has never been observed for visible optics (as it requires $\mu\neq1$). Figure 4a shows a reflection image of one of our structures in polarized green light with the electric field along the *y*-axis. In this polarization, as discussed earlier, the QR is not excited. Accordingly, the patterned medium is seen as a darker area on top of a glass substrate. On the other hand, if the same structure is illuminated in the perpendicular polarisation, which does excite the quadrupole mode, the intensity contrast disappears so that the sample is no longer visible (Fig.4b). The disappearance is due to the matching between impedances of air ($Z_{air}$=1) and the patterned structure ($Z_s$=1). The latter exhibits both $\varepsilon$ and $\mu$ equal to 0.73+0.13i for $\lambda$=540nm, as found from the analysis of the sample's spectra. At the same time, the difference in optical properties of the two media can still be visualised because they have different refraction indices $n$ ($n_s$=0.73+0.13i $\neq$ $n_{air}$=1). This leads to a phase difference acquired by the reflected light, which was converted to an intensity contrast by using the standard phase contrast technique (Fig.4c).

In summary, we have proved the concept of an artificial medium with large magnetic response at optical frequencies, including regions with negative both $\mu$ and $\varepsilon$. This is a principal step forward from the previously demonstrated magnetism at microwave frequencies towards left-handed visible optics.

Figure Captions.

Fig. 1. Nanofabricated medium with negative permeability at optical frequencies. (a) Scanning electron micrograph (viewed at an angle) of an array of Au nanopillars. (b,c) – Numerical simulation of the distribution of electric currents inside a pair of such pillars for symmetric and antisymmetric resonances, respectively. There exist two other symmetric and antisymmetric modes with dipoles oscillating along *x*- and *y*-axes. For perfectly cylindrical pillars, the symmetry forbids the excitation of antisymmetric modes for any direction of incident light. However, due to the slightly conical shape of the pillars, which breaks the centre symmetry, light incident along *z*-axis with the electric field along *x*-axis can excite the antisymmetric mode shown in (c). The non-cylindrical shape was intentionally introduced in our design through a choice of microfabrication procedures and is a result of Au evaporation through an opening in a resist mask, which becomes increasingly narrower during evaporation. Light incident along *z*-axis used in our experiments also excites the symmetric modes in *x-y* plane (not shown).

Fig. 2. Reflection spectra for different arrays of Au nanopillars. (a) #1 - control sample with individual (non-interacting) pillars; #2 and #3 – samples with interacting pillars (*s*=200nm and 140nm, respectively). Periodicity *D* is 600nm for all three samples. The spectra are offset for clarity. The insets show micrographs of the corresponding samples. The solid curves are best fits to theoretical dependences with $\lambda_s$=667nm, $\Delta\lambda_s$=150nm for sample #1, $\lambda_s$=680nm, $\Delta\lambda_s$=160nm, $\lambda_a$=510nm, $\Delta\lambda_a$=55nm for #2 and $\lambda_s$=690nm, $\Delta\lambda_s$=170nm, $\lambda_a$=500nm, $\Delta\lambda_a$=57nm for #3. Blue arrows indicate positions of dipole and quadrupole resonances. The effective thickness was measured by ellipsometry and was close to the mass thickness ($\approx$10nm for all three samples), in agreement with numerical simulations [24,25]. (b) Zoom in the spectrum of sample #3 near the "green" plasmon resonance. (c) Dependence of the position

of plasmon resonances on distance *s* between neighbouring pillars. Symbols are experimental data for the symmetric (dipoles oscillate along *x*-axis) and antisymmetric (dipoles oscillate along *z*-axis) resonances; solid lines are theoretical fit.

Fig. 3. Excitation anisotropy and enhancement of magnetic response. (a) The reflection spectra measured for two polarizations of the normal incident light with the electric field along *x*-axis (green symbols) and *y*-axis (red). The circles are for sample #4 (*s*=140nm, *D*=400nm); squares are for the same sample covered with a thin layer of glycerine (*n*=1.47). The insets show optical photographs of sample #4 (size 0.3x0.3mm$^2$) observed at the normal angle of white light incidence for the two light polarizations. White stripes beneath the optical photographs show the colour of a non-patterned gold film of the same thickness (note that the plasmon resonance in a continuous film cannot be excited by perpendicular incident light, for symmetry considerations). We emphasize that the observed polychroism is not an interference or form-birefringence effect but appears due to the sensitivity of the excited plasmon resonances to the incident light polarization. (b) Reflection spectra for monochromatic normal incident light as a function of its polarisation angle $\theta$ plotted in the polar co-ordinates for sample #4. The polar distance corresponds to $\lambda$ varying from 400 to 700nm; the polar angle $\theta$ is counted as shown in the insert to the right. The colour scale corresponds to the reflection coefficient varying between 7 and 25%.

Fig. 4. Impedance matching. Optical photograph of sample #5 illuminated by green light with the electric field along *y*- (a) and *x*-axis (b). The sample becomes invisible (no intensity contrast) in the case of (b). *d*=120nm, *h*=80nm, *s*=200nm, *D*=600nm; pattern over an area of 0.2x0.2mm$^2$. (c) Optical photograph of the same sample illuminated by the same green light as in (b) but viewed using a phase contrast microscopy. Note that the intensity contrast can also disappear due to interference effects but in our case this possibility is ruled out because of a small effective optical thickness of our media and no evidence for the interference at other wavelengths. The phase matching requires matching of both real and imaginary parts of the impedance, which places rather stringent conditions on relation between the strengths and wavelengths of magnetic and dipole resonances. These conditions were met for the particular structure, which exhibited $f_s$=0.15, $\lambda_s$=650nm, $\Delta\lambda_s$=120nm, $f_a$=0.034, $\lambda_a$=570nm, $\Delta\lambda_a$=30nm.

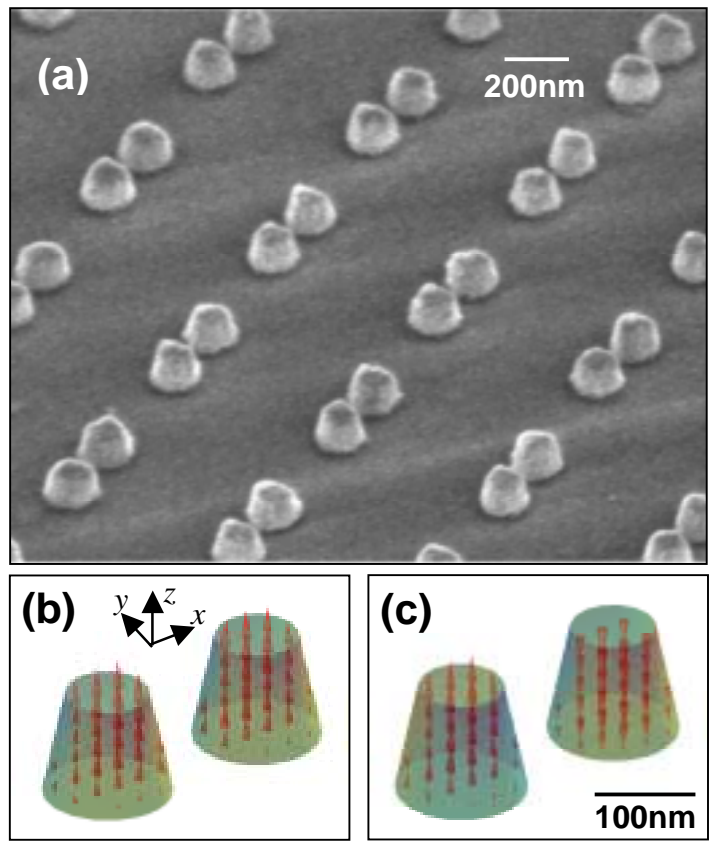

**Figure 1.**

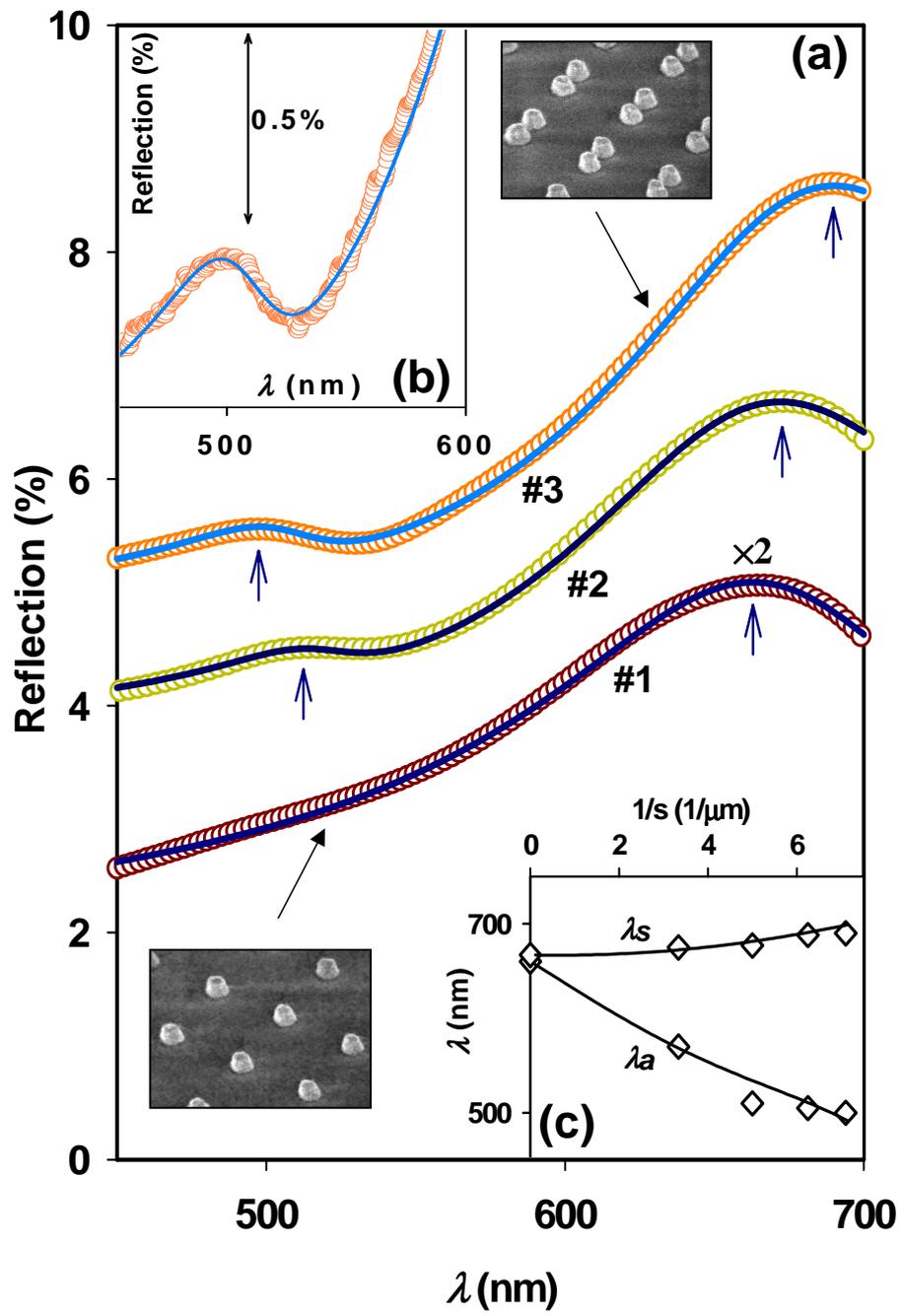

Figure 2.

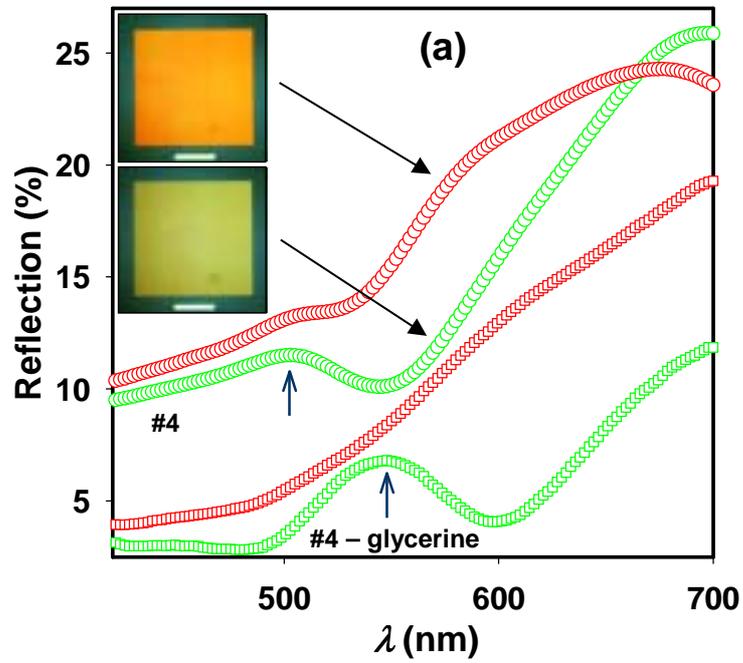

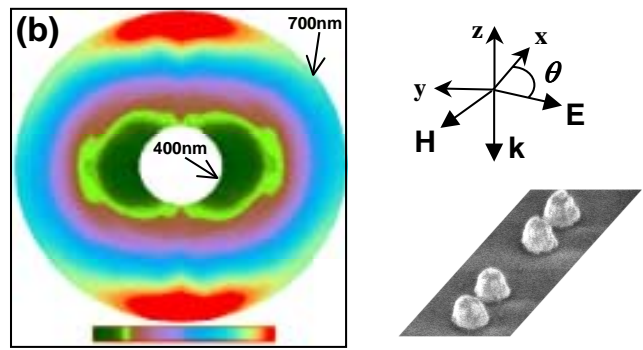

**Figure 3.**

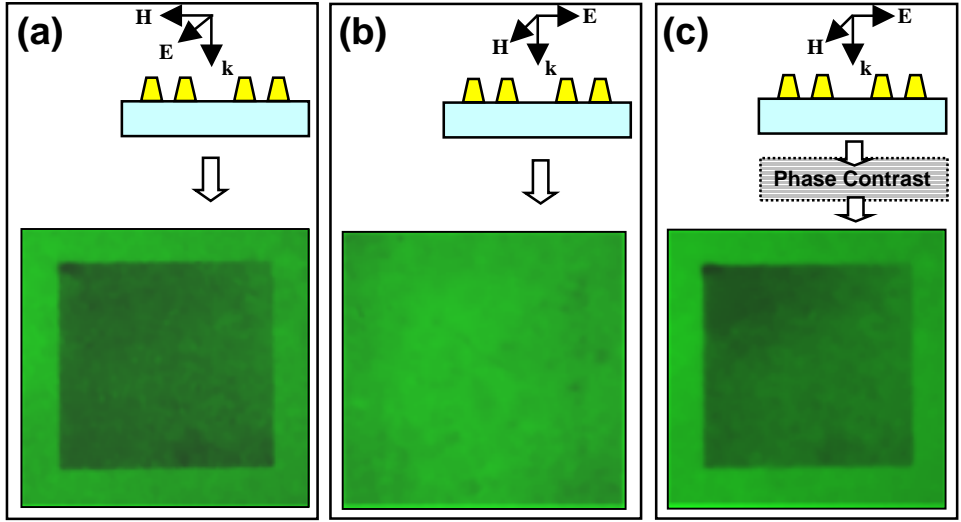

**Figure 4.**